\documentclass[useAMS,usegraphicx,usenatbib]{mn2e}
\usepackage{lscape}

\newcommand{\mnras}{MNRAS}

\title[Filtering out activity-related variations]{Filtering out activity-related variations from radial velocities in a search for low-mass planets}

\author[M. Tuomi et al.]{M. Tuomi$^{1,2,3}$\thanks{E-mail: \texttt{mikko.tuomi@utu.fi}; \texttt{m.tuomi@herts.ac.uk}},
G. Anglada-Escud\'e$^{4,5}$, 
J. S. Jenkins$^{2}$,
H. R. A. Jones$^{1}$
\\
$^1$University of Hertfordshire, Centre for Astrophysics Research, Science and Technology Research Institute, College Lane, AL10\\ 9AB, Hatfield, UK\\
$^2$Departamento de Astronom\'ia, Universidad de Chile, Camino del Observatorio 1515, Las Condes, Santiago, Chile\\
$^3$University of Turku, Tuorla Observatory, Department of Physics and Astronomy, V\"ais\"al\"antie 20, FI-21500, Piikki\"o, Finland\\
$^4$Institut f\"ur Astrophysik, Georg-August Universitat G\"ottingen, Friedrich-Hund-Platz 1, 37077 G\"ottingen, Germany\\
$^5$School of Physics and Astronomy, Queen Mary University of London, 327 Mile End Road, E1 4NS, London, United Kingdom\\
}

\begin{document}

\date{Accepted {XX.XX.2014}. Received {XX.XX.2014}; in original form {XX.XX.2014}}

\pagerange{\pageref{firstpage} -- \pageref{lastpage}} \pubyear{2014}

\maketitle

\label{firstpage}

\begin{abstract}
  Variations related to stellar activity and correlated noise can prevent the detections of low-amplitude signals in radial velocity data if not accounted for. This can be seen as the greatest obstacle in detecting Earth-like planets orbiting nearby stars with Doppler spectroscopy regardless of developments in instrumentation and rapidly accumulating amounts of data. We use a statistical model that is not sensitive to aperiodic and/or quasiperiodic variability of stellar origin. We demonstrate the performance of our model by re-analysing the radial velocities of the moderately active star CoRoT-7 ($\log R_{\rm HK} = -4.61$) with a transiting planet with a transiting planet whose Doppler signal has proven rather difficult to detect. We find that the signal of the transiting planet can be robustly detected together with signals of two other planet candidates. Our results suggest that rotation periods of moderately active stars can be filtered out of the radial velocity noise, which enables the detections of low-mass planets orbiting such stars.
\end{abstract}



\section{Introduction}

Activity-related noise caused by rotation, active features, and irregularities of stellar surfaces \citep[e.g.][]{santos2010,dumusque2011,dumusque2012,hatzes2013,tuomi2013b} provide currently the greatest challenges in improving the precision of radial velocity surveys to such an extent that detections of Earth-like planets become possible. Despite being largely induced by rotation, the nature of stellar noise related to activity gives rise to aperiodic and/or quasiperiodic variations that can mimic periodic Doppler signals of Keplerian origin and thus prevent the detections of low-amplitude planetary signals by littering the posterior densities (or periodograms) with spurious signals that can be mistaken for genuine signals of planetary origin. In order to increase the sensitivity of radial velocity surveys, it is necessary to develop statistical models \citep[e.g. for correlated noise:][]{baluev2013,tuomi2013b} accounting for these irregular variations. Furthermore, it should only be accepted that a weak planetary signal is detected if it is necessary in describing the radial velocity variations and that these variations cannot be explained by stochastic variability that might be connected to measures of stellar activity and/or intrinsic correlations in the data \citep[e.g.][]{boisse2011,pont2011}.

CoRoT-7 is a moderately active G9 ($T_{\rm eff} = 5250\pm60$; [Fe/H] $=  0.12\pm0.06$ dex; $M_{\star} = 0.91\pm0.03$ M$_{\oplus}$; $\log R_{\rm HK} = -4.612$) dwarf star \citep{queloz2009,bruntt2010} that has a planetary system orbiting it, including one transiting planet corresponding to a valuable benchmark planet for estimating the compositions and structures of such super-Earths \citep{leger2009,queloz2009}. However, determination of the mass of this planet based on high-precision radial velocities has proved difficult \citep[e.g.][]{queloz2009,hatzes2010,boisse2011,hatzes2011,pont2011} due to the difficulties in detecting the signal in the precense of at least equally significant variations related to stellar activity and rotation \citep{queloz2009,ferrazmello2011,hatzes2011} and because modelling approaches based on different assumptions and approaches -- pre-whitening or high-pass filtering corresponding to the removal of periodicities estimated to correspond to activity-related phenomena \citep{queloz2009,ferrazmello2011} or modelling starspot-induced radial velocity variations \citep{lanza2010,pont2011} -- have provided a wide range of estimates that are consistent at best but sometimes very diverse \citep[see the estimates listed in][]{hatzes2011}.

In this work we obtain the HARPS spectra of CoRoT-7 from the European Southern Observatory archive, process them with the \emph{Template-Enhanced Radial velocity Re-analysis Application} \citep[HARPS-TERRA;][]{anglada2012c}, and obtain velocity products together with activity indicators. We aim to filter out the activity-related variations from the CoRoT-7 radial velocity data with our noise model. Spurious quasiperiodic signals related to the stellar rotation of roughly 23 d \citep{lanza2010} corresponding to the ``dominant frequency in the power spectrum'' according to \citet{queloz2009}, have been affecting the obtained solutions to the radial velocities. Our goal is the detection of the signal of the transiting planet without any prior knowledge of its orbital period and, more generally, to be able to distinguish between activity-related variability and genuine Doppler signals of transiting planets.

\section{Statistical modelling}

Stellar activity can contribute considerably to the variations in precision radial velocity data \citep[e.g.][]{lagrange2010,santos2010,boisse2011,dumusque2011}. It is thus necessary to account for these variations with the statistical model in order to detect low-amplitude signals of planetary origin. Our key point is that activity variations are not strictly periodic and comprise quasi- and/or aperiodic variability that is unlikely to be modelled accurately by using Keplerian periodicities or sinusoids. This is the case even when such variations appear as periodicities in the commonly used Lomb-Scargle periodograms \citep{lomb1976,scargle1982,cumming2004}, although such an interpretation might be tempting in practice \citep[e.g.][]{queloz2009,dumusque2012,hatzes2013} in the context of so-called ``pre-whitening'' that aims at removing periodogram powers related to activity and/or rotation by subtracting the corresponding sinusoids from the data. Since such variations cannot be expected to be strictly sinusoidal, this procedure gives rise to harmonics of the subtracted signals that need to be subtracted as well in order to decrease the contribution of such features to the data. Such an approach cannot be considered very satisfactory in practice because such a subtraction of sinusoids that are not orthonormal in the case of unevenly sampled time series causes biases to the genuine signals in the data \citep[e.g.][]{pont2011}, but also, because it requires prior knowledge of the nature of the signals seen in the periodograms, e.g. information from transit observations, such that activity-related spurious signals are subtracted but planetary signals are left intact \citep{queloz2009,ferrazmello2011}. We note that more sophisticated periodogram-based methods have been proposed as well, including the generalised log-likelihood periodograms \citep[e.g.][and references therein]{anglada2012b,baluev2013} that evaluate the significances of signals based on likelihood-ratio tests and can be applied to arbitrary noise models, and the minimum mean squared error technique for determining the optimal number of signals in a data set \citep{jenkins2014} that also provides means for testing whether the signals are stationary in time.

Instead, we adopt the approach of \citet{tuomi2013b} and assume that a large fraction of activity-induced variability can in fact be filtered out from the velocities by accounting for intrinsic correlations in the data. In practice, this means that the radial velocity noise is not white but has a significant red-noise component even for stars that are known to be rather inactive \citep{baluev2013,tuomi2013a,tuomi2013b,feroz2014}. However, such models for correlated noise might not be sufficient in explaining the variations in radial velocities of moderately active stars such as CoRoT-7. Well-known measures of stellar activity, obtained from the same spectra as the velocities themselves, can be useful \citep[e.g.][]{pont2011} to estimate how much the stellar activity contributes to the radial velocity variability. In particular, we consider using the line bisector span (BIS) that respond to spots and other surface inhomogeneities co-rotating on the stellar surface \citep{saar1998,desort2007}, line full-width at the half-maximum (FWHM), and the $S$-index based on the CaII H\&K lines as reasonable measures of activity \citep[e.g.][]{boisse2011}.

The above considerations suggest, as a first-order approximation, a model for the radial velocities that can be written \citep[see also][]{tuomi2013b} as
\begin{eqnarray}\label{eq:model}
  m_{i} = \gamma + \dot{\gamma}t_{i} + f_{k}(t_{i}) + \epsilon_{i} + \sum_{j=1}^{p} \phi_{j} \exp \Bigg\{ \frac{t_{i-j} - t_{i}}{\tau} \Bigg\} \epsilon_{i} \nonumber\\
  + \sum_{j=1}^{q} c_{j} \xi_{j,i} ,
\end{eqnarray}
where the measurement $m_{i}$ made at time $t_{i}$ is modelled by using the reference velocity $\gamma$; a linear trend quantified by using the parameter $\dot{\gamma}$ that is can be present in the radial velocities of any given target due to the possible existence of a long-period stellar and/or substellar companion; a white-noise component $\epsilon_{i}$ that we assume to have a Gaussian density with a zero mean and a variance of $\sigma_{i}^{2} + \sigma_{\rm J}^{2}$, where $\sigma_{i}$ represents the estimated instrument noise and $\sigma_{J}$ is a free parameter of the model; a correlated noise component that we describe according to a $p$th order moving average (MA) model with an exponential smoothing in the time-scale of $\tau$ \citep{baluev2013,tuomi2013b}; and linear correlations with the activity indices $\xi_{j,i}$ quantified by using the parameters $c_{j}$.

Although it is possible that the linear correlations between velocities and activity indices described above are not sufficient as the relationship between those two cannot be expected to be linear (or in fact to follow any simple functional description), we adopt such a description as a first-order approximation. While we do not expect this model to be optimal, if velocity variations are indeed
impacted by activity this will be better than modelling the velocity variations without such a component. Moreover, the MA($p$) term might not be an optimal description of the correlated noise either. For instance, \citet{baluev2013} and \citet{feroz2014} describe the noise such that they set all $\phi_{j}$ equal and take the summation over all $j = 1, ..., N$, $j \neq i$. We do not consider this a realistical description. First, such a model corresponds to a violation of causality as the measurements made after the $i$th one contribute to its noise, although this might be justifiable as simply a statistical description without such an interpretation. Second, the assumption that the parameter $\phi_{j}$ is the same for all $j = 1, ..., N$ might be too restrictive. Furthermore, using such models requires inversions of $N \times N$ matrices, where $N$ is the number of measurements, which becomes computationally heavy even for moderately high $N$. We choose an MA(1) model that does not increase the computational requirements and appears to be a reasonably good noise model in practice \citep{tuomi2014,tuomi2014b} and because increasing the number of MA components did not improve the model. Moreover, as we measure activity-related information based on three activity indicators, namely BIS, FWHM, and $S$-indices, we set $p = 1$ and $q = 3$ in Eq. (\ref{eq:model}).

The model described in Eq. \ref{eq:model} implies a likelihood model for the measurements that can be calculated for the $i$th measurement based on the obtained values $\epsilon_{j}$, where $j = i-1, ..., i-p$. We used the priors as discussed in \citep{tuomi2013c} such that the eccentricity prior was defined as $\pi(e) \propto \mathcal{N}(0, 0.1^{2})$ and the jitter prior as $\pi(\sigma_{\rm J}) \propto \mathcal{N}(0, \sigma_{\sigma}^{2})$, where $\mu_{\sigma} = \sigma_{\sigma}^{2} =$ 2 ms$^{-1}$. See \citet{tuomi2013c} and \citet{anglada2013} for discussion and justification.

\section{DRAM samplings}

We analyse the radial velocity data sets by applying the delayed-rejection adaptive-Metropolis (DRAM) algorithm \citep{haario2006} that is a generalisation of the adaptive-Metropolis algorithm \citep{haario2001} based on the Metropolis-Hastings algorithm \citep{metropolis1953,hastings1970}.

The basics of the DRAM algorithm are as follows. We first draw a proposal ($\theta_{1}$) from a proposal density $q_{1}(\theta_{0}, \theta_{1})$. This density is a function of the current state of the chain $\theta_{0}$ (typically centered at it). The proposal is then accepted with the probability
\begin{equation}\label{eq:acceptance}
  \alpha_{1}(\theta_{0}, \theta_{1}) = \min \Bigg\{1, \frac{\pi(\theta_{1} | m) q_{1}(\theta_{1}, \theta_{0})}{\pi(\theta_{0} | m) q_{1}(\theta_{0}, \theta_{1})} \Bigg\} ,
\end{equation}
where $\pi(\theta | m)$ is the posterior density given the data $m$. If the proposal density is a symmetric one, as is the case when applying a multivariate Gaussian density, the acceptance probability depends only on the posterior ratio of the two vectors because $q_{1}(\theta_{0}, \theta_{1}) = q_{1}(\theta_{1}, \theta_{0})$. Assuming that the new vector $\theta_{1}$ is rejected, another vector $\theta_{2}$ can then be proposed by using another proposal density $q_{2}(\theta_{0}, \theta_{1}, \theta_{2})$ \citep{haario2006}. In addition to the current state $\theta_{0}$ and the newly proposed vector $\theta_{2}$, this new proposal density can depend on the rejected vector $\theta_{1}$ as well. The new proposal is then accepted with probability
\begin{eqnarray}\label{eq:acceptance2}
  \alpha_{2}(\theta_{0}, \theta_{1}, \theta_{2}) = \min \Bigg\{1, \frac{\pi(\theta_{2} | m) q_{1}(\theta_{2}, \theta_{1}) q_{2}(\theta_{2}, \theta_{1}, \theta_{0})}{\pi(\theta_{0} | m) q_{1}(\theta_{0}, \theta_{1}) q_{2}(\theta_{0}, \theta_{1}, \theta_{2})} \nonumber\\
  \times \frac{[1 - \alpha_{1}(\theta_{2}, \theta_{1})]}{[1 - \alpha_{1}(\theta_{0}, \theta_{1})]} \Bigg\} .
\end{eqnarray}
This process can be repeated for an arbitrary number of rejected proposals in the above manner. The equation for the general case, i.e. $n$th acceptance probability, can be found in \citet{haario2006}.

Analyses of radial velocity time series in a search for periodic signals are typically complicated because the period parameter (its logarithm in our analyses) has a highly multimodal likelihood function and thus posterior density. Therefore, it is necessary to perform the DRAM samplings in such a way that the chains visit all the modes in the posterior regularly and that the highest maxima are identified robustly in the samplings. For this reason, we choose the proposal densities such that $q_{1}$ is a multivariate Gaussian density with a covariance matrix $C_{1}$. The second proposal is then a multivariate Gaussian as well with a covariance of $C_{2}$ that is obtained by multiplying the row and the column corresponding to the period parameter of the signal by a factor of $\delta_{i} = 0.3$ such that the width of the proposal density is decreased in the dimension of the period to enable the chains to find the narrow probability maxima in the period space. We allowed three delayed-rejection steps before finally rejecting a proposed vector. Such a practice has been proposed by \citet{haario2006} albeit in a different context.

When searching for periodic signals in the data, we used tempered samplings such that a posterior raised to the power of $\beta \in (0,1)$ was used instead of the original posterior. This ensured, by decreasing the relative heights of the modes in the posterior, that the corresponding Markov chains visited all the relevant areas in the period space. This is important because the posterior density is typically highly multimodal in such examples of analyses of radial velocity data \citep[e.g.][]{tuomi2014,tuomi2014b}. DRAM samplings were not used to obtain point and uncertainty estimates for the model parameters but only for ensuring that there were no additional significant maxima in the period space. For parameter estimation purposes, we set $\beta = 1$ to draw samples from the original posterior density with the adaptive-Metropolis algorithm.

\section{Searching for signals in HARPS data of CoRoT-7}

The CoRoT-7 Doppler data have been obtained over two rather distinct observing runs between 2454527 - 2454885 JD and between 2455939 - 2455965 JD. We call these simply the first ($N = 95$ after removing few measurements corresponding to FWHM outliers) and second observing run ($N = 71$), respectively. These velocities are tabulated in the appendix. Because there is a gap of almost three years between these two observing runs, we believe that it cannot be expected that the noise properties, including all possible correlations as modelled according to Eq. (\ref{eq:model}), are equivalent in these runs due to differences in stellar activity, e.g. the evolution of spot patterns, and the designs of the observations in terms of e.g. cadence and exposure time. Therefore, we assume that the noise parameters of these two observing runs are independent of one another. Even if they are not strictly independent, we believe that treating them as free parameters for each run, as one would for data from different instruments, enables us to determine whether there are indeed differences that should be accounted for and to obtain a better statistical model than when making the much stronger assumption that the noise properties were not evolving as a function of time, which is unlikely to be the case \citep[e.g.][]{queloz2009}. Due to this choice, our model contains eight independent and two common parameters for the two observing runs, which yields a baseline model with 18 parameters when there are no Keplerian signals in the model.

\begin{figure*}
\center
\includegraphics[angle=270, width=0.32\textwidth, clip]{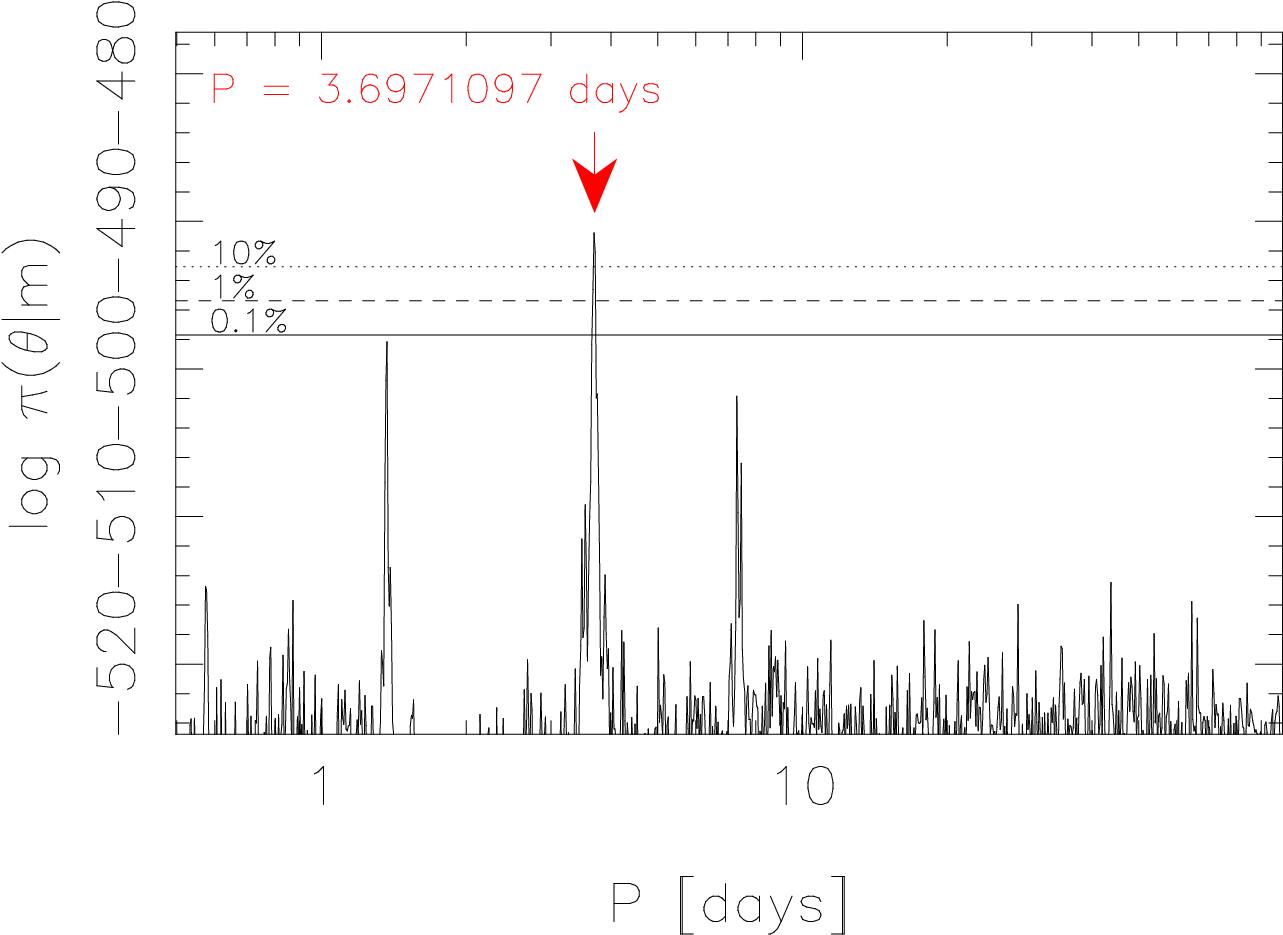}
\includegraphics[angle=270, width=0.32\textwidth, clip]{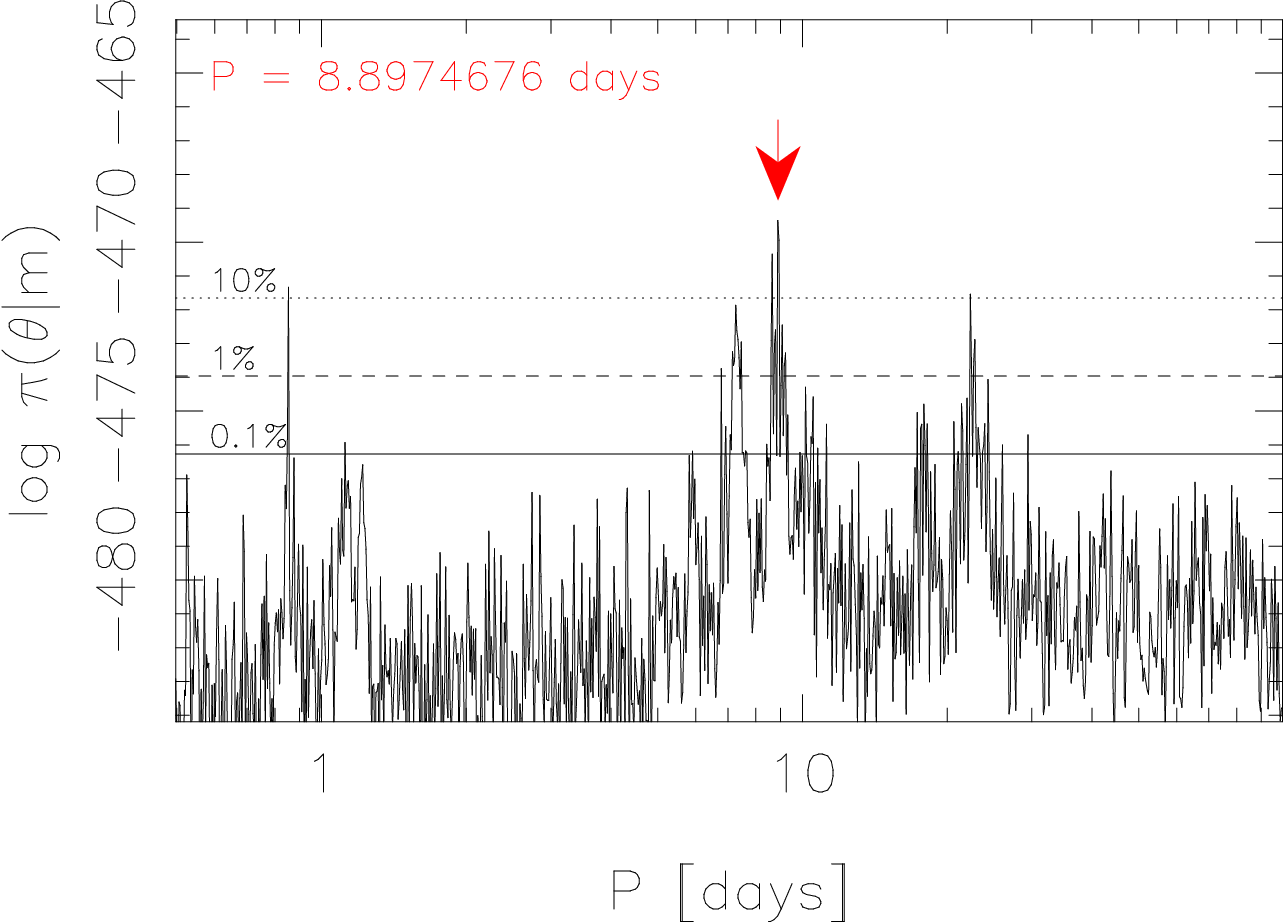}
\includegraphics[angle=270, width=0.32\textwidth, clip]{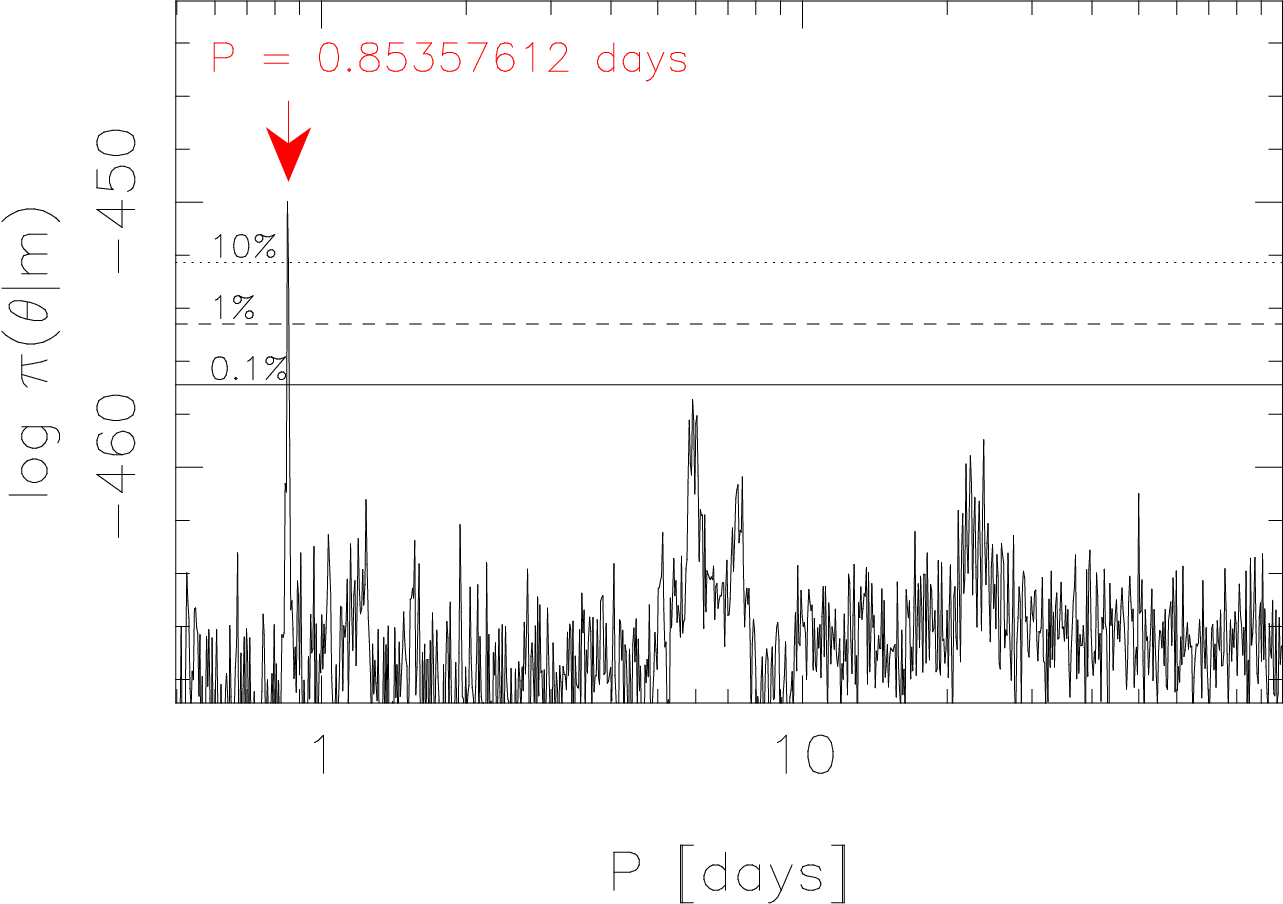}
\caption{Estimated posterior densities of the three signals in the CoRoT-7 radial velocity data as functions of signal period based on tempered DRAM samplings of the one, two, and three-Keplerian models. The maximum values of each sampling are denoted by red arrows and the horizontal lines denote the 10\% (dotted), 1\% (dashed), and 0.1\% (solid) thresholds with respect to the maxima. The period space has been limited to an interval between 0.5 d and 100 d.}\label{fig:dram_searches}
\end{figure*}

\begin{figure*}
\center
\includegraphics[angle=270, width=0.32\textwidth, clip]{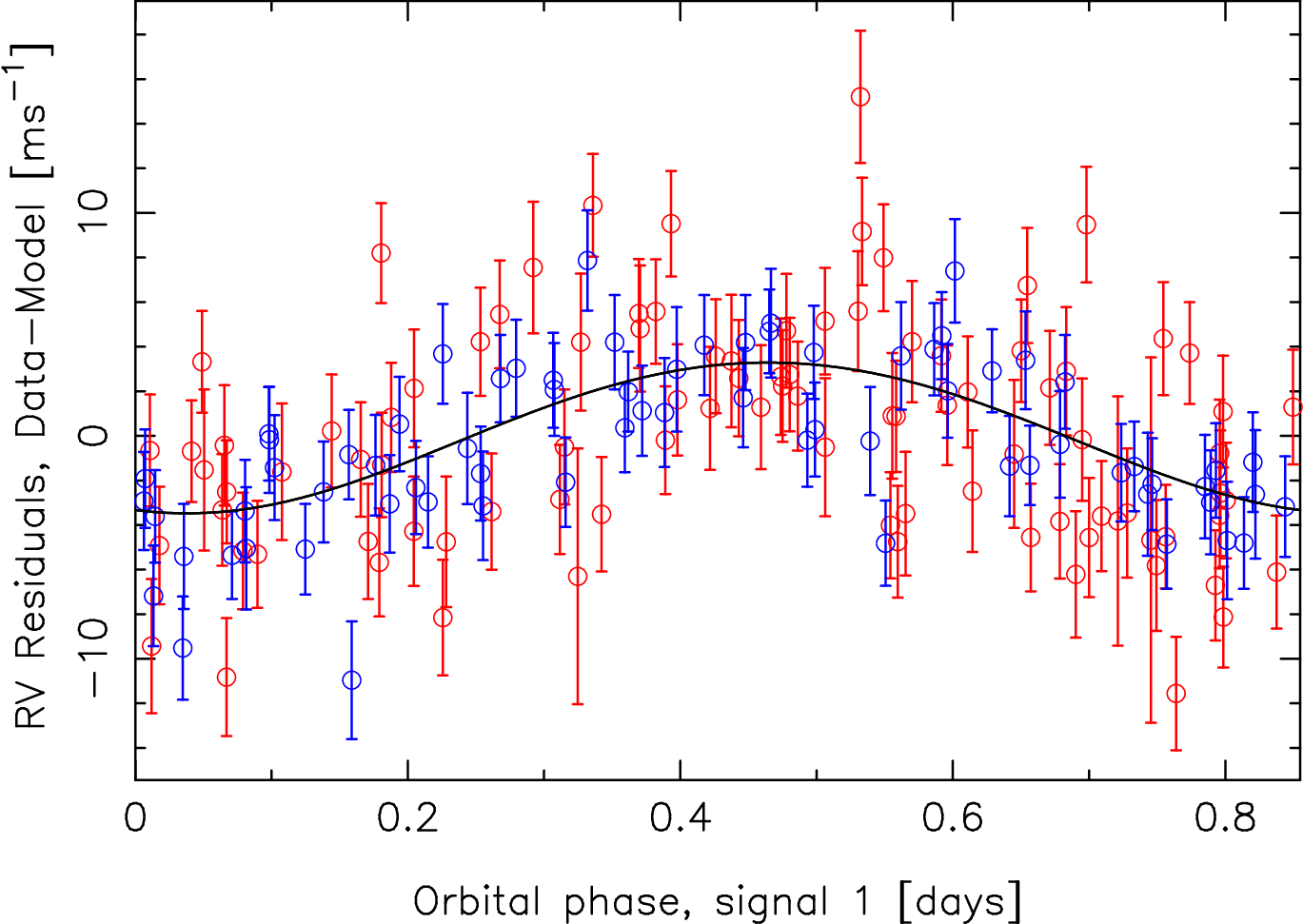}
\includegraphics[angle=270, width=0.32\textwidth, clip]{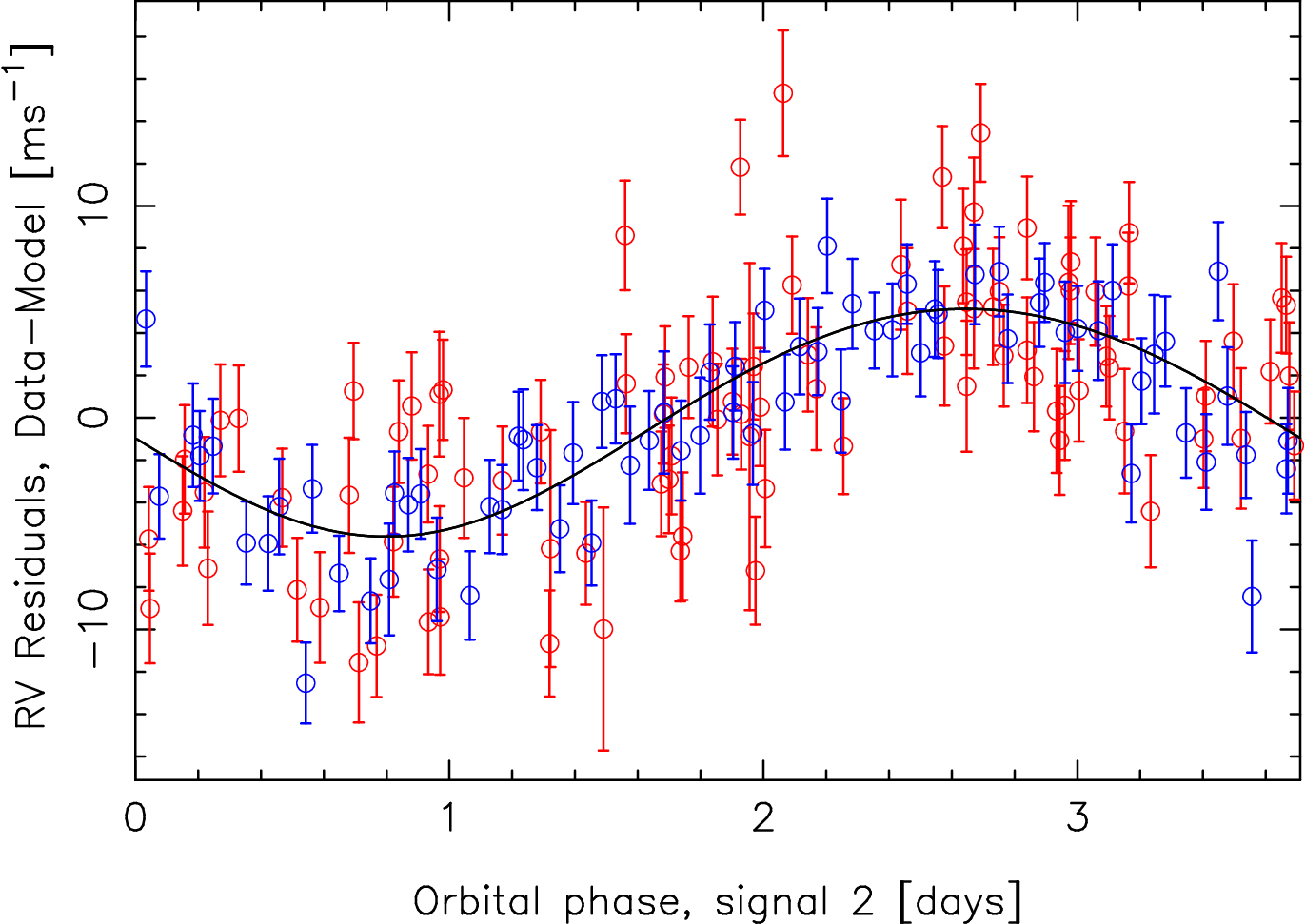}
\includegraphics[angle=270, width=0.32\textwidth, clip]{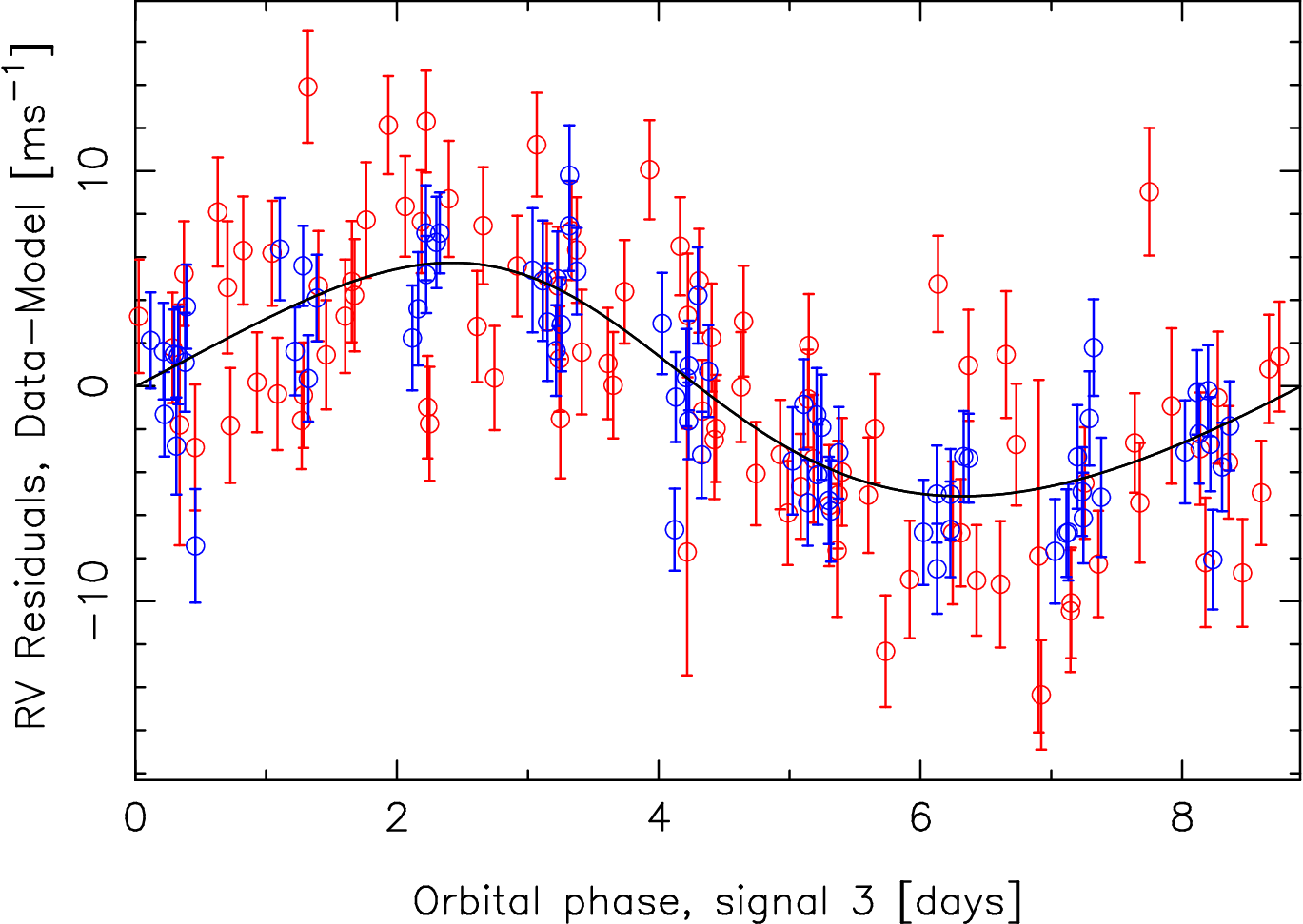}

\includegraphics[angle=270, width=0.32\textwidth, clip]{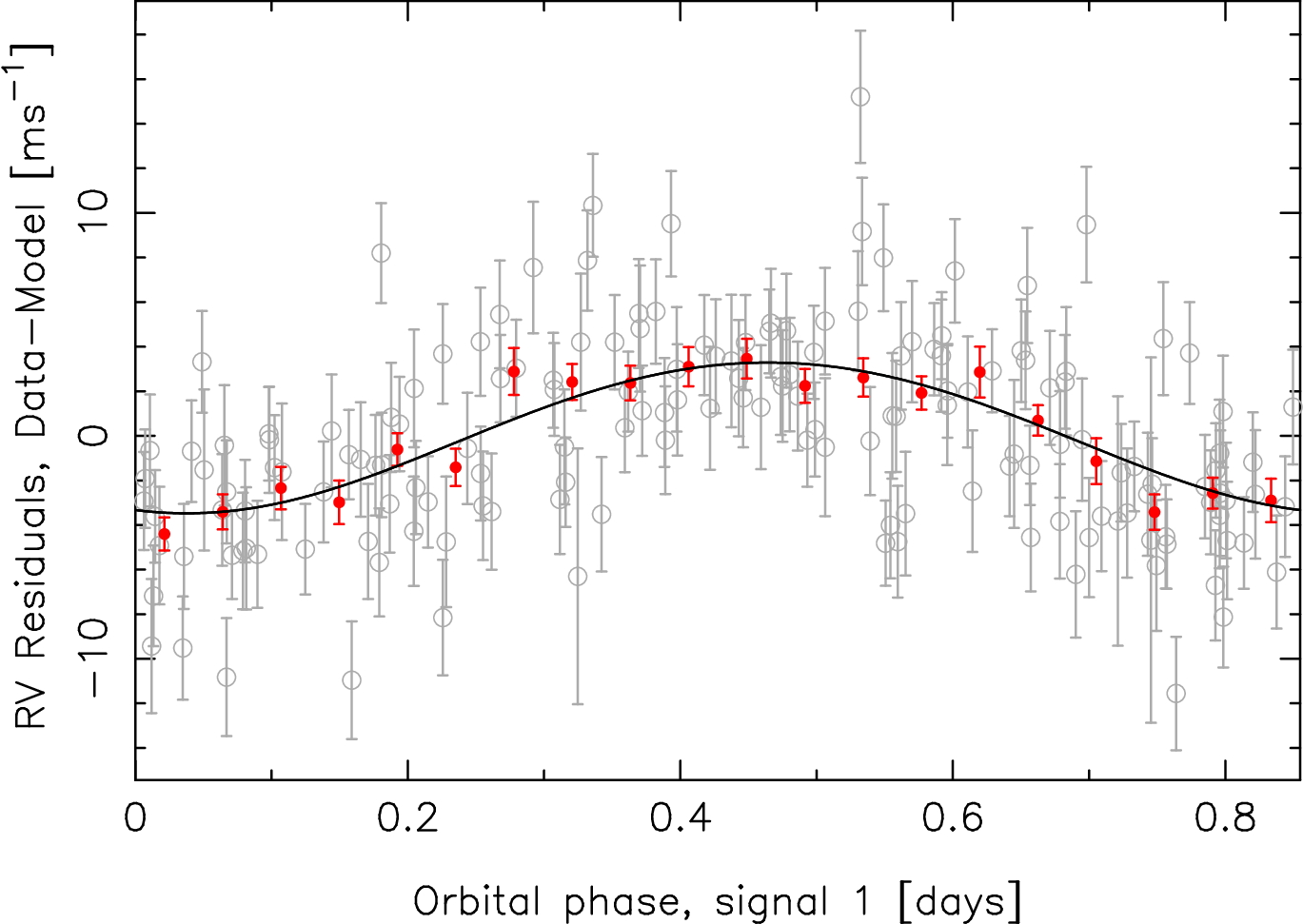}
\includegraphics[angle=270, width=0.32\textwidth, clip]{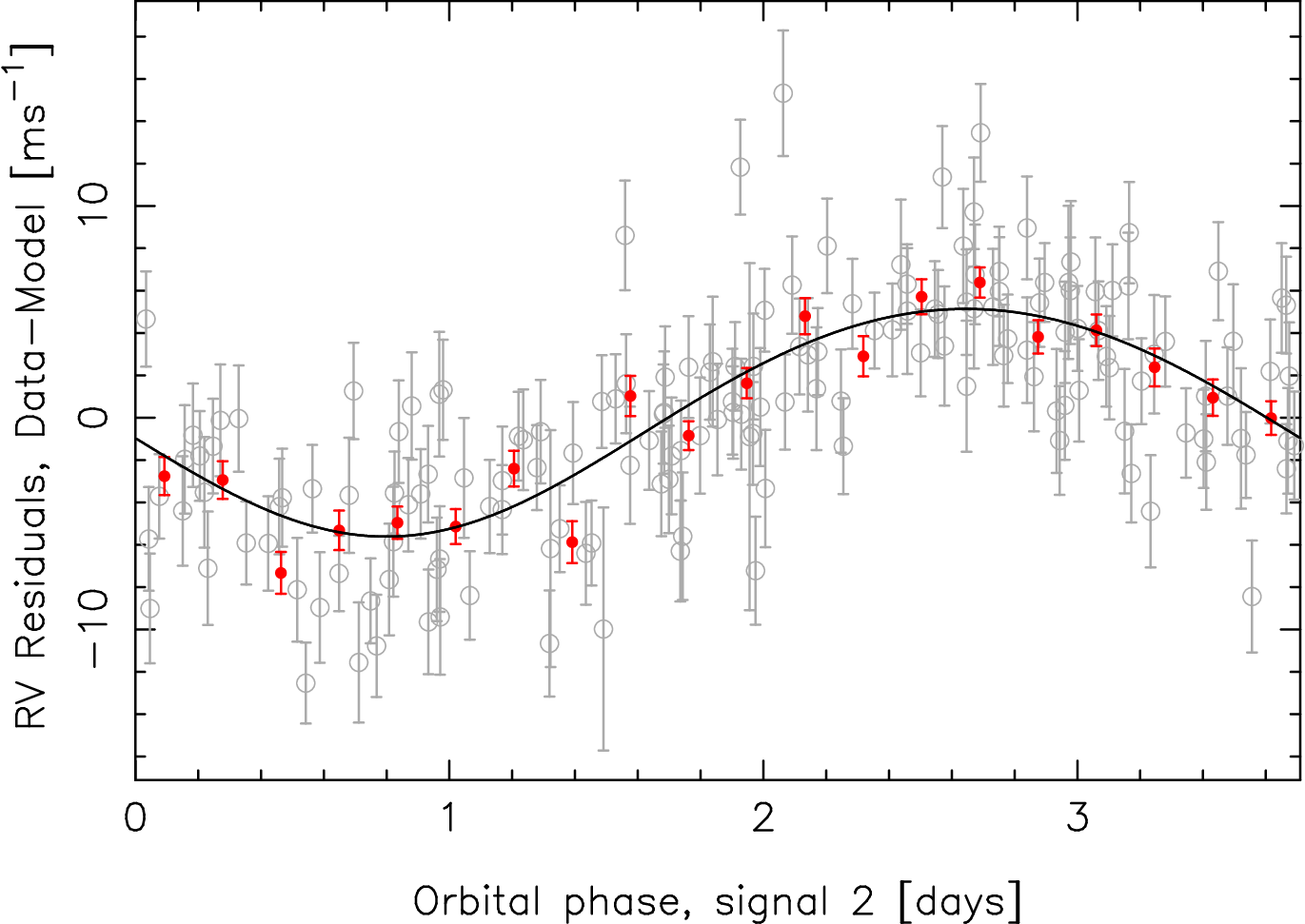}
\includegraphics[angle=270, width=0.32\textwidth, clip]{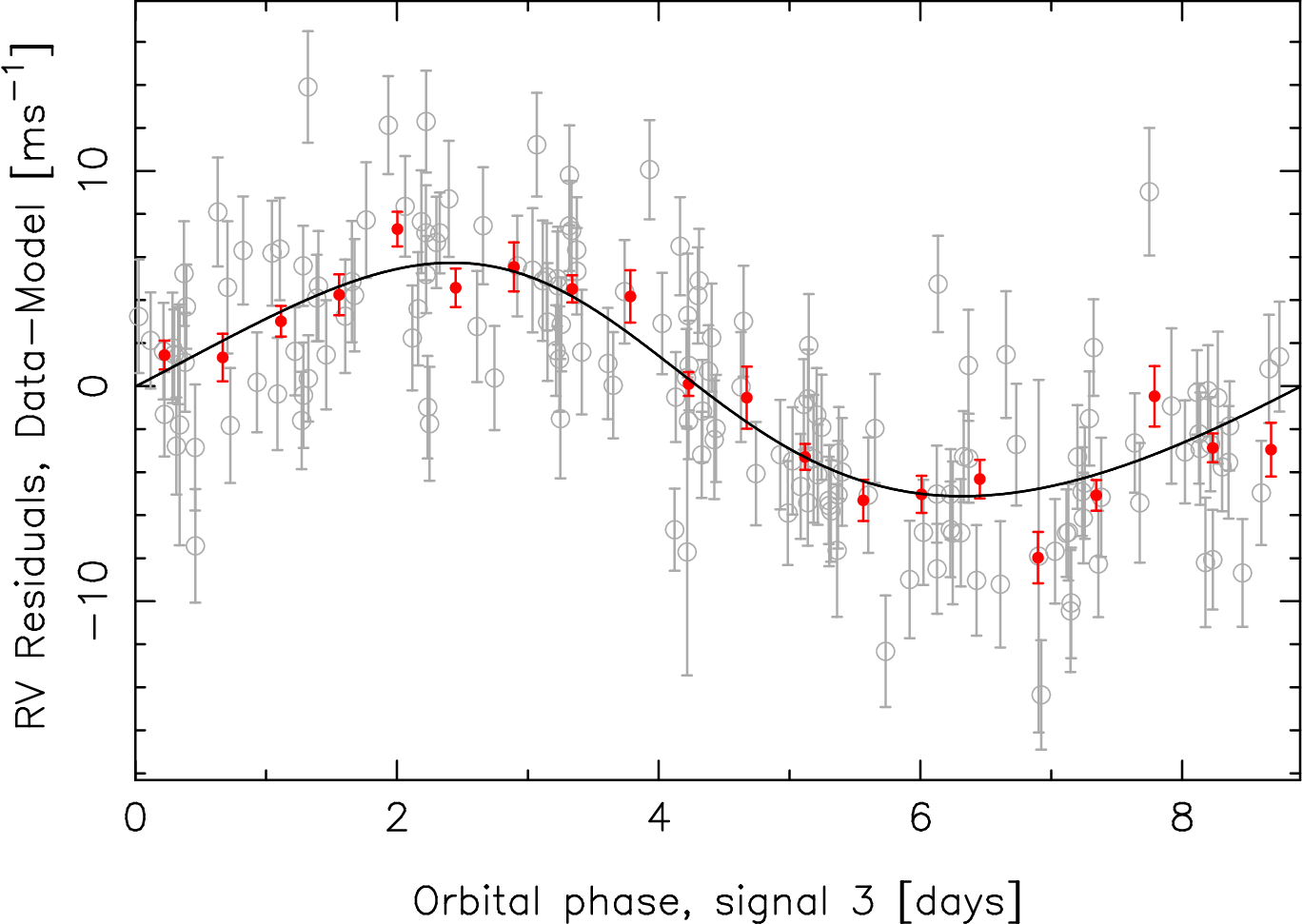}
\caption{Top: phase-folded signals of the three candidate planets orbiting CoRoT-7. Red and blue points denote the first and second HARPS observing run, respectively. The top panels denote all the data and the bottom panels show weighted means (red filled circles) when the phases of the signals have been divided into 20 bins.}\label{fig:signals}
\end{figure*}

We started the analyses of the HARPS velocity data by performing tempered samplings of a one-Keplerian model. These samplings enabled us to identify a signal at 3.7 d as a unique probability maximum in the period space together with local maxima corresponding to its double and a daily alias at a period of 1.4 d (Fig. \ref{fig:dram_searches}, right hand panel). Similar samplings, albeit with parameter $\beta$ set to values closer to unity, of a two-Keplerian model yielded a maximum at a period of 9.0 d for the second signal (Fig. \ref{fig:dram_searches} middle panel) together with local maxima at 0.86 d and 22 d. After this, additional samplings revealed a unique third signal in the data at a period of 0.86 d (Fig. \ref{fig:dram_searches}, right hand panel) without considerable local maxima. This sequence of signal detections was consistent with the signal detection criteria of \citet{tuomi2012} as the log-Bayesian evidences were increased at these steps by 43.0, 13.2, and 19.0. These log-Bayesian evidences correspond to model probabilities that increase by factors of $4.9 \times 10^{18}$, $5.5 \times 10^{5}$, and $1.9 \times 10^{8}$, respectively, when assuming equal prior probabilities for each model. The phase-folded signals are shown in Fig. \ref{fig:signals}. We did not detect any additional signals in the data.

The observed rotation period of the star of 23 d that has been detected in the radial velocities in earlier studies \citep{queloz2009} did not correspond to the global maximum for any of the models used in our analyses. Although it can be seen as a local maximum in the DRAM search for the second most significant signal (Fig. \ref{fig:dram_searches}, middle panel), it did not correspond to a significant solution even when we started additional samplings in its vicinity. Furthermore, we did not observe significant signals at that period even after accounting for the three candidate planets orbiting the star. This indicates that the statistical model (Eq. \ref{eq:model}) accounts for the aperiodic and quasiperiodic variations corresponding to the stellar activity sufficiently well and makes it unnecessary to model the rotation period (and/or its harmonics) as additional signals in the data \citep{queloz2009,boisse2011}. We interpret this result as a suggestion that the noise model is an adequate description of the activity-related variations and, more generally, that it might be possible to remove such variability from radial velocity data of at most moderately active stars. This could therefore help in the detection of low amplitude radial velocity variations corresponding to low-mass planets around such stars.

We did not find significant correlations between the radial velocities and BIS values or the $S$-indices. However, the FWHM appears to be connected to the velocity variations as we obtain $c_{\rm FWHM} =$ 0.094 [0.016, 0.198] for the first observing run and $c_{\rm FWHM} =$ 0.065 [-0.013, 0.192] for the second when using the maximum a posteriori (MAP) estimates and the 99\% credibility intervals. The former correlation is significantly present in the data considering that the 99\% interval does not overlap with zero but the latter is significant in this sense with respect to only a 95\% credibility interval.

The excess noise, as measured by obtaining estimates for the parameters $\sigma_{\rm J}$, is likely different between the two observing runs. We obtained estimates of 3.89 [3.07, 4.97] ms$^{-1}$ and 1.88 [1.42, 3.21] ms$^{-1}$ for the first and second observing runs, respectively.  Similarly, the first observing run contained significant intrinsic correlations because we obtained $\phi_{1} =$ 0.88 [0.68, 1], whereas the same parameter was consistent with zero for the second observing run. The time-scale ($\tau$) of these intrinsic correlations was found to be roughly 6 d \citep[see also][]{baluev2013,tuomi2013b}. Not accounting for these differences might weight the first observing run too much and the second one too little and lead to biased results.

Finally, we obtained estimates for the proposed system of three planets and show them in Table \ref{tab:parameters}.

\begin{table*}
\caption{Parameter estimates corresponding to the three-Keplerian solution to the HARPS-TERRA radial velocities of CoRoT-7 in terms of MAP estimates and the 99\% credibility intervals. The credibility intervals of semi-major axes and minimum masses have been estimated by assuming an uncertainty in the stellar mass of 10\%.}\label{tab:parameters}
\begin{minipage}{\textwidth}
\begin{center}
\begin{tabular}{lccc}
\hline \hline
Parameter & CoRoT-7 b\footnote{Transiting planet whose inclination is known to be 80.1$^{\circ}$ \citep{leger2009}.} & CoRoT-7 c & CoRoT-7 d \\
\hline
$P$ (d) & 0.85487 [0.85479, 0.85496] & 3.7094 [3.7084, 3.7103] & 8.8999 [8.8938, 8.9081] \\
$K$ (ms$^{-1}$) & 3.41 [1.86, 4.96] & 5.30 [4.06, 6.42] & 5.16 [3.32, 7.00] \\
$e$ & 0.14 [0, 0.23] & 0.01 [0, 0.16] & 0.07 [0, 0.23] \\
$\omega$ (rad) & 3.7 [0, 2$\pi$] & 5.8 [0, 2$\pi$] & 1.7 [0, 2$\pi$] \\
$M_{0}$ (rad) & 1.7 [0, 2$\pi$] & 2.5 [0, 2$\pi$] & 3.8 [0, 2$\pi$] \\
$m_{p} \sin i$ (M$_{\oplus}$) & 4.8 [2.4, 7.1] & 11.8 [8.4, 15.9] & 15.4 [9.4, 22.8] \\
$a$ (AU) & 0.0172 [0.0153, 0.0187] & 0.0458 [0.0409, 0.0498] & 0.0820 [0.0733, 0.0892] \\
\hline \hline
\end{tabular}
\end{center}
\end{minipage}
\end{table*}

We note that according to our tests, samplings with the DRAM algorithm proved more robust than the corresponding samplings with the adaptive-Metropolis algorithm. We obtained the results by generating Markov chains with few $10^{7}$ members with DRAM, whereas we could obtain consistent results with the adaptive-Metropolis algorithm with roughly ten times longer chains. This is caused by the fact that the DRAM algorithm enables the chains to visit narrow high-probability regions in the parameter space more easily. This is because the sampling algorithms that are capable of exploring the posterior locally are more efficient in exploring the period space of radial velocity signals in practice due to a multimodality of the corresponding probability densities.

\section{Discussion}

We have presented an analysis of the HARPS-TERRA velocities of CoRoT-7 that has been proposed to host a system of two planetary companion, possibly three, out of which the innermost one with an orbital period of roughly 0.85 days is transiting. According to our results, the three signals can be detected based on the HARPS-TERRA radial velocities by performing DRAM searches for posterior maxima in the period space without any prior knowledge of the period of the transiting planet \citep[e.g.][]{queloz2009} that has the weakest radial velocity signal. Furthermore, with the statistical model accounting for both intrinsic correlations in the data and linear correlations with activity indices (Eq. \ref{eq:model}), we did not observe the stellar rotation period at 23 days \citep{queloz2009} as a global maximum with models containing any number of Keplerian signals. This suggests that for moderately active stars such as CoRoT-7 with $\log R_{\rm HK} = -4.612$, such modelling of correlations automatically filters out activity-induced signals and variability related to the stellar rotation. We note that although we used the BIS, FWHM, and $S$-index values as indicators of stellar activity, other measures of such activity should be considered as well \citep[e.g.][]{barnes2014,santos2014}. Choosing the most informative indices and the exact functional form for the correlations requires model comparisons with a handful of benchmark targets and is a subject of future work together with the choice of the red noise model. However, our simple approach appears to work well for the CoRoT-7 data.

Modelling the activity-related variations based on observed photometric variability is another approach that has been attempted \citep{lanza2010}. However, apart from photometric detections of transiting planets and stellar rotation periods, it is not clear how photometric data that is more often than not obtained at different epochs than the Doppler data, such as is the case with CoRoT-7 \citep{pont2011}, can be connected to velocity variations in a useful and trustworthy manner. For this reason, as suggested by \citet{pont2011}, the information available in the activity data obtained from the same spectra as the radial velocities should be accounted for when modelling the relationship between stellar activity and radial velocities.

With an estimated mass of $4.8_{-2.4}^{+2.3}$ M$_{\oplus}$, and adopting a radius of 1.69$\pm$0.09 R$_{\oplus}$ \citep{leger2009}, the average density of this transiting planet is $5.6^{+4.2}_{-3.1}$ gcm$^{-3}$ that does not constrain the planetary composition very accurately. Most of the uncertainty in this estimate arises from the uncertainty in estimating the planetary mass better than by a factor of roughly 50\%. The above mass estimate appears to be consistent with all the earlier estimates due to its uncertainty \citep[see][]{hatzes2011} and underlines the importance of improving the statistical models used to analyse Doppler spectroscopy data in order to achieve greater sensitivity to low-amplitude signals and better constraints for the planetary masses.

\section*{Acknowledgements}

The authors acknowledge the significant efforts of the HARPS-ESO team in improving the instrument and its data reduction pipelines.



\begin{thebibliography}{100}\small
\bibitem[\protect\astroncite{Anglada-Escud\'e \& Butler}{2012}]{anglada2012c} Anglada-Escud\'e, G. \& Butler, R. P. 2012, ApJS, 200, 15
\bibitem[\protect\astroncite{Anglada-Escud\'e \& Tuomi}{2012}]{anglada2012b} Anglada-Escud\'e, G. \& Tuomi, M. 2012, A\&A, 548, A58
\bibitem[\protect\astroncite{Anglada-Escud\'e et al.}{2013}]{anglada2013} Anglada-Escud\'e, G., Tuomi, M., Gerlach, E., et al. 2013, A\&A, 556, A126
\bibitem[\protect\astroncite{Baluev}{2013}]{baluev2013} Baluev, R. V. 2013, MNRAS, 429, 2052
\bibitem[\protect\astroncite{Barnes et al.}{2014}]{barnes2014} Barnes, J. R., Jenkins, J. S., Jones, H. R. A., et al. 2014, MNRAS, 439, 3094
\bibitem[\protect\astroncite{Boisse et al.}{2011}]{boisse2011} Boisse, I., Bouchy, F., H\'ebrard, G., et al. 2011, A\&A, 528, A4
\bibitem[\protect\astroncite{Bruntt et al.}{2010}]{bruntt2010} Bruntt, H., Deleuil, M., Fridlund, M., et al. 2010, A\&A, 519, A51
\bibitem[\protect\astroncite{Cumming}{2004}]{cumming2004} Cumming, A. 2004, \mnras, 354, 1165
\bibitem[\protect\astroncite{Desort et al.}{2007}]{desort2007} Desort, M., Lagrange, A.-M., Galland, F., et al. 2007, A\&A, 473, 983
\bibitem[\protect\astroncite{Dumusque et al.}{2011}]{dumusque2011} Dumusque, X., Lovis, C., S\'egransan, D., et al. 2011, A\&A, 535, A55
\bibitem[\protect\astroncite{Dumusque et al.}{2012}]{dumusque2012} Dumusque, X., Pepe, F., Lovis, C., et al. 2012, Nature, 491, 207 
\bibitem[\protect\astroncite{Feroz et al.}{2014}]{feroz2014} Feroz, F. \& Hobson, M. P. 2014, MNRAS, 437, 3540
\bibitem[\protect\astroncite{Ferraz-Mello et al.}{2011}]{ferrazmello2011} Ferraz-Mello, S., Tadeu dos Santos, M., Beaug\'e, C., et al. 2011, A\&A, 531, A161
\bibitem[\protect\astroncite{Haario et al.}{2001}]{haario2001} Haario, H., Saksman, E., \& Tamminen, J. 2001, Bernoulli, 7, 223
\bibitem[\protect\astroncite{Haario et al.}{2006}]{haario2006} Haario, H., Laine, M., Mira, A., and Saksman, E. 2006, Statistics and Computing, 16, 339
\bibitem[\protect\astroncite{Hastings}{1970}]{hastings1970} Hastings, W. 1970, Biometrika 57, 97
\bibitem[\protect\astroncite{Hatzes et al.}{2010}]{hatzes2010} Hatzes, A. P., Dvorak, R., Wuchterl, G., et al. 2010, A\&A, 520, A93
\bibitem[\protect\astroncite{Hatzes et al.}{2011}]{hatzes2011} Hatzes, A. P., Fridlund, M., Nachmani, G., et al. 2011, ApJ, 743, 75
\bibitem[\protect\astroncite{Hatzes}{2013}]{hatzes2013} Hatzes, A. P. 2013, ApJ, 770, 133
\bibitem[\protect\astroncite{Jenkins et al.}{2014}]{jenkins2014} Jenkins, J. S., Becerra Yoma, N., Rojo, P., et al. 2014, arXiv:1403.7646
\bibitem[\protect\astroncite{Kipping}{2013}]{kipping2013} Kipping, D. M. 2013, MNRAS, 434, L51
\bibitem[\protect\astroncite{Lagrange et al.}{2010}]{lagrange2010} Lagrange, A.-M., Desort, M., Meunier, N. 2010, A\&A, 512, 38
\bibitem[\protect\astroncite{Lanza et al.}{2010}]{lanza2010} Lanza, A. F., Bonomo, A. S., Moutou, C., et al. 2010, A\&A, 520, A53
\bibitem[\protect\astroncite{L\'eger et al.}{2009}]{leger2009} L\'eger, A., Rouan, D., Schneider, J., et al. 2009, A\&A, 506, 287
\bibitem[\protect\astroncite{Lomb}{1976}]{lomb1976} Lomb, N. R. 1976, Astrophys. Space Sci., 39, 447
\bibitem[\protect\astroncite{Metropolis et al.}{1953}]{metropolis1953} Metropolis, N., Rosenbluth, A. W., Rosenbluth, M. N., et al. 1953, J. Chem. Phys., 21, 1087
\bibitem[\protect\astroncite{Pepe et al.}{2011}]{pepe2011} Pepe, F., Lovis, C., S\'egransan, D., et al. 2011, A\&A, 534, A58
\bibitem[\protect\astroncite{Pont et al.}{2011}]{pont2011} Pont, F., Aigrain, S., \& Zucker, S. 2011, MNRAS, 411, 1953
\bibitem[\protect\astroncite{Queloz et al.}{2009}]{queloz2009} Queloz, D., Bouchy, F., Moutou, C., et al. 2009, A\&A, 506, 303
\bibitem[\protect\astroncite{Saar et al.}{1998}]{saar1998} Saar, S. H., Butler, R. P., \& Marcy, G. W. 1998, ApJ, 498, 153
\bibitem[\protect\astroncite{Santos et al.}{2010}]{santos2010} Santos, N. C., Gomez da Silva, J., Lovis, C., \& Melo, C. 2010, A\&A, 511, A54
\bibitem[\protect\astroncite{Santos et al.}{2014}]{santos2014} Santos, N. C., Mortier, A., Faria, J. P., et al. 2014, arXiv:1404.6135
\bibitem[\protect\astroncite{Scargle}{1982}]{scargle1982} Scargle, J. D. 1982, ApJ, 263, 835
\bibitem[\protect\astroncite{Tuomi}{2012}]{tuomi2012} Tuomi, M. 2012, A\&A, 543, A52
\bibitem[\protect\astroncite{Tuomi}{2014}]{tuomi2014} Tuomi, M. 2014, MNRAS, 440, L1
\bibitem[\protect\astroncite{Tuomi \& Anglada-Escud\'e}{2013}]{tuomi2013c} Tuomi, M. \& Anglada-Escud\'e 2013, A\&A, 556, A111
\bibitem[\protect\astroncite{Tuomi et al.}{2013a}]{tuomi2013a} Tuomi, M., Anglada-Escud\'e, G., Gerlach, E., et al. 2013a A\&A, 549, A48
\bibitem[\protect\astroncite{Tuomi et al.}{2013b}]{tuomi2013b} Tuomi, M., Jones, H. R. A., Jenkins, J. S., et al. 2013b, A\&A, 551, A79
\bibitem[\protect\astroncite{Tuomi et al.}{2014}]{tuomi2014b} Tuomi, M., Jones, H. R. A., Barnes, J. R., et al. 2014, arXiv:1403.0430
\end{thebibliography}

\appendix

\section{HARPS-TERRA radial velocities and activity indices}

\begin{table*}
\caption{HARPS-TERRA radial velocities and activity indices}\label{tab:data}
\centering
\begin{tabular}{@{}lccccc}
\hline \hline
Time & Velocity & Unc. & BIS & FWHM & $S$-index \\

(JD) & (ms$^{-1}$) & (ms$^{-1}$) & (ms$^{-1}$) & (ms$^{-1}$) \\
\hline \hline
2454527.5439 &  -4.99 & 1.54 & -10.77 & 7065.46 & 0.313 \\
2454530.6043 & -10.22 & 1.54 &  -8.95 & 7063.34 & 0.292 \\
2454550.5035 &   9.15 & 2.01 & -16.08 & 7072.77 & 0.348 \\
2454775.8191 &   5.82 & 1.92 &  -2.20 & 7106.35 & 0.360 \\
2454776.7585 &   5.06 & 2.38 &  -7.85 & 7118.73 & 0.341 \\
2454777.7617 &   1.04 & 1.67 &   1.00 & 7099.14 & 0.354 \\
2454778.7546 & -10.32 & 1.48 &  -8.62 & 7100.19 & 0.348 \\
2454779.7507 &  -1.98 & 1.71 &  -8.39 & 7112.18 & 0.328 \\
2454780.7548 &   2.33 & 2.16 &  -6.81 & 7099.48 & 0.349 \\
2454789.8299 &   4.31 & 1.59 &  -8.79 & 7110.22 & 0.338 \\
2454790.8030 &  10.30 & 1.75 & -17.08 & 7111.44 & 0.332 \\
2454791.8110 &  14.95 & 1.66 & -11.83 & 7126.43 & 0.360 \\
2454792.8100 &   7.85 & 1.66 & -15.37 & 7137.91 & 0.351 \\
2454793.8277 &   4.09 & 1.61 &  -4.38 & 7134.45 & 0.354 \\
2454794.8006 &   7.42 & 1.82 &  -6.20 & 7137.91 & 0.343 \\
2454795.8090 &   0.43 & 2.17 &  -9.36 & 7143.53 & 0.355 \\
2454796.8365 &  -7.05 & 1.20 & -11.89 & 7128.05 & 0.370 \\
2454797.8034 &  -5.06 & 1.41 &  -3.17 & 7126.94 & 0.346 \\
2454798.8061 &  11.32 & 2.36 &   8.36 & 7124.53 & 0.330 \\
2454799.7770 &  11.94 & 1.71 &  -3.08 & 7112.73 & 0.345 \\
2454799.8645 &  11.32 & 1.83 & -11.02 & 7083.73 & 0.340 \\
2454800.7591 &   6.71 & 1.87 &  11.47 & 7095.61 & 0.332 \\
2454800.8462 &   2.59 & 1.39 &   2.40 & 7111.80 & 0.373 \\
2454801.7530 &  -0.63 & 1.48 &   3.10 & 7095.20 & 0.342 \\
2454801.8396 &   0.44 & 1.37 &  -2.41 & 7099.11 & 0.334 \\
2454802.7484 &   5.11 & 1.65 &  -9.66 & 7092.29 & 0.337 \\
2454802.8433 &   2.19 & 1.37 &  -6.45 & 7090.82 & 0.336 \\
2454803.7526 &  -2.47 & 1.59 &  -8.16 & 7092.55 & 0.330 \\
2454803.8342 &  -8.96 & 1.68 & -20.38 & 7090.10 & 0.329 \\
2454804.7555 &  -4.30 & 2.19 &  -9.92 & 7090.11 & 0.337 \\
2454804.8339 &   0.09 & 2.03 & -15.24 & 7096.25 & 0.323 \\
2454805.7775 &   7.79 & 1.96 & -21.53 & 7100.44 & 0.322 \\
2454805.8505 &  20.54 & 2.21 & -20.96 & 7099.34 & 0.333 \\
2454806.7647 &  22.53 & 1.56 & -14.13 & 7104.13 & 0.321 \\
2454806.8444 &  22.47 & 1.62 & -14.31 & 7105.15 & 0.341 \\
2454807.7281 &  11.59 & 1.80 & -14.30 & 7111.29 & 0.342 \\
2454807.8264 &  12.17 & 1.55 & -16.46 & 7111.45 & 0.336 \\
2454825.7368 &   4.76 & 1.22 & -19.30 & 7094.43 & 0.339 \\
2454826.7395 &   6.04 & 1.14 & -21.99 & 7101.59 & 0.331 \\
2454827.7262 &  14.49 & 1.26 & -24.33 & 7117.61 & 0.341 \\
2454828.7361 &  28.97 & 1.19 & -22.56 & 7142.23 & 0.318 \\
2454829.7354 &  18.68 & 1.61 & -21.42 & 7150.01 & 0.349 \\
2454830.7244 &   3.46 & 1.88 &  -6.63 & 7149.14 & 0.342 \\
2454831.7103 &   4.41 & 7.95 &  -9.99 & 7136.02 & 0.384 \\
2454831.7288 &  -6.17 & 1.61 & -11.17 & 7128.55 & 0.381 \\
2454832.7252 &   2.99 & 3.03 & -11.82 & 7156.55 & 0.385 \\
2454833.7342 &   1.29 & 1.75 &  -8.47 & 7133.01 & 0.342 \\
2454834.7539 &  14.48 & 1.44 & -13.89 & 7159.14 & 0.363 \\
2454847.5968 &   2.98 & 1.39 &  -8.49 & 7068.19 & 0.356 \\
2454847.6939 &   2.74 & 1.43 &  -4.75 & 7081.79 & 0.367 \\
2454847.7568 &   6.95 & 1.35 &   4.27 & 7089.08 & 0.349 \\
2454852.5986 &  -0.58 & 1.70 & -10.95 & 7096.83 & 0.342 \\
2454853.5740 &  11.56 & 1.26 &  -4.00 & 7100.32 & 0.348 \\
2454853.6981 &  14.54 & 1.36 &  -8.31 & 7085.68 & 0.368 \\
2454853.7461 &  10.68 & 1.33 & -11.85 & 7111.19 & 0.380 \\
2454854.5803 &  20.57 & 1.38 & -15.43 & 7097.61 & 0.370 \\
2454854.6578 &  18.36 & 1.51 &  -3.17 & 7096.64 & 0.358 \\
2454854.7417 &  16.39 & 1.90 & -12.22 & 7094.49 & 0.362 \\
2454855.6755 &  14.10 & 1.15 & -16.20 & 7134.77 & 0.349 \\
2454856.6528 &   6.08 & 1.14 & -18.14 & 7142.73 & 0.351 \\
2454857.6468 &  22.99 & 1.06 & -25.00 & 7149.07 & 0.364 \\
2454858.6635 &  21.94 & 1.64 & -24.23 & 7159.29 & 0.365 \\
2454859.6499 &  15.22 & 1.71 & -12.67 & 7164.50 & 0.354 \\
2454860.7522 &  15.08 & 5.24 & -12.12 & 7182.90 & 0.479 \\
2454861.6845 &  12.74 & 1.10 &  -0.73 & 7166.63 & 0.371 \\
2454862.6627 &   5.76 & 1.77 &  -8.08 & 7160.32 & 0.376 \\
2454863.6550 &  -2.77 & 1.45 &  -2.86 & 7164.14 & 0.352 \\
2454864.6302 &  -6.02 & 5.39 & -15.50 & 7173.11 & 0.332 \\
2454865.5978 &  -1.95 & 2.22 &  -0.17 & 7137.06 & 0.348 \\
2454865.7159 &  -4.64 & 2.01 &  -0.39 & 7125.86 & 0.346 \\
2454866.6602 & -10.46 & 2.67 &   1.74 & 7132.21 & 0.345 \\
2454867.5604 & -20.67 & 2.04 &  -3.95 & 7134.08 & 0.330 \\
2454867.6706 & -22.22 & 1.68 &  -1.31 & 7128.70 & 0.350 \\
2454868.5918 & -18.27 & 2.27 & -14.16 & 7115.99 & 0.359 \\
2454868.6876 & -14.15 & 2.34 & -17.00 & 7121.74 & 0.335 \\
2454869.6002 &  -4.38 & 1.64 &  -8.57 & 7106.96 & 0.335 \\
2454869.6880 &   4.07 & 1.41 & -11.69 & 7119.14 & 0.337 \\
2454870.6013 &  -1.28 & 1.45 &  -9.39 & 7116.62 & 0.336 \\
2454870.7157 &   1.81 & 1.62 & -14.23 & 7126.08 & 0.333 \\
2454871.5378 &   6.79 & 1.30 & -15.59 & 7105.56 & 0.347 \\
2454872.5648 &   7.60 & 1.95 &  -6.79 & 7107.10 & 0.360 \\
2454872.6496 &   8.49 & 1.17 & -11.84 & 7100.89 & 0.331 \\
2454872.7273 &   6.93 & 2.12 & -18.82 & 7116.92 & 0.344 \\
2454873.5375 &   5.94 & 2.09 & -13.92 & 7096.67 & 0.309 \\
2454873.6501 &   4.13 & 1.32 & -21.63 & 7112.66 & 0.336 \\
2454873.7201 &   4.60 & 1.55 & -11.08 & 7120.03 & 0.297 \\
2454879.5366 &  17.03 & 1.68 &  -9.52 & 7109.84 & 0.325 \\
2454879.6761 &  11.16 & 1.59 & -14.52 & 7099.59 & 0.313 \\
2454880.6137 &  20.68 & 1.84 &  -6.51 & 7116.50 & 0.320 \\
2454881.5911 &  18.66 & 1.46 & -19.24 & 7124.07 & 0.351 \\
2454882.5257 &  14.94 & 1.40 & -17.40 & 7132.51 & 0.346 \\
2454882.6560 &  15.88 & 1.52 & -14.35 & 7138.26 & 0.367 \\
2454883.5880 &  16.16 & 1.53 & -15.09 & 7133.97 & 0.346 \\
2454884.5263 &  14.24 & 1.52 &  -5.80 & 7132.32 & 0.359 \\
2454884.6464 &   6.51 & 1.64 &  -2.24 & 7135.16 & 0.332 \\
2455939.6995 & -12.46 & 2.24 & -24.56 & 7035.96 & 0.285 \\
2455939.7602 &  -9.43 & 1.71 & -11.54 & 7049.23 & 0.281 \\
2455940.5750 &  -3.50 & 2.52 & -22.34 & 7054.56 & 0.296 \\
2455940.6893 &  -7.68 & 2.36 &  -7.91 & 7044.84 & 0.260 \\
2455940.7946 &  -3.45 & 1.68 & -22.10 & 7058.45 & 0.274 \\
2455941.5649 &  -3.47 & 1.90 & -30.27 & 7040.22 & 0.256 \\
2455941.6687 &  -3.66 & 1.56 & -16.52 & 7049.63 & 0.270 \\
2455941.7702 &  -0.37 & 1.55 & -16.26 & 7051.96 & 0.265 \\
2455942.5614 & -10.48 & 2.07 & -31.50 & 7043.00 & 0.304 \\
2455942.6770 & -11.84 & 1.42 & -16.10 & 7046.99 & 0.287 \\
2455942.7841 &  -9.19 & 2.00 &  -2.90 & 7047.67 & 0.283 \\
2455943.5609 & -16.52 & 2.01 & -22.32 & 7043.54 & 0.279 \\
2455943.6657 & -18.63 & 1.55 & -13.32 & 7041.16 & 0.272 \\
2455943.7687 & -16.17 & 1.58 & -12.82 & 7044.50 & 0.266 \\
2455944.5667 & -10.34 & 1.99 & -17.02 & 7043.41 & 0.275 \\
2455944.6691 & -11.92 & 1.76 & -10.45 & 7032.89 & 0.281 \\
2455944.7737 & -11.24 & 1.52 & -18.40 & 7036.27 & 0.283 \\
2455945.5610 &  -5.73 & 1.94 & -27.75 & 7049.05 & 0.270 \\
2455945.6674 &  -8.10 & 1.86 & -27.44 & 7043.99 & 0.272 \\
2455945.7721 & -16.75 & 1.85 & -12.75 & 7037.58 & 0.266 \\
2455946.5574 & -12.87 & 1.75 & -36.10 & 7039.95 & 0.258 \\
2455946.6631 & -16.39 & 1.38 & -24.64 & 7039.15 & 0.272 \\
2455946.7684 & -12.66 & 1.78 & -19.35 & 7051.15 & 0.261 \\
2455947.5453 &  -8.75 & 1.93 & -15.26 & 7043.29 & 0.267 \\
2455947.6617 &  -8.77 & 1.50 & -25.89 & 7049.17 & 0.274 \\
2455947.7628 &  -7.41 & 1.45 & -21.82 & 7058.24 & 0.258 \\
2455948.5571 &  -0.84 & 2.00 & -24.28 & 7053.76 & 0.271 \\
2455948.6636 &   5.34 & 1.13 & -18.51 & 7062.51 & 0.271 \\
2455948.7672 &   8.43 & 1.26 & -16.30 & 7062.14 & 0.260 \\
2455949.5541 &   2.66 & 2.42 & -11.06 & 7054.19 & 0.234 \\
2455949.6555 &   0.21 & 1.58 & -19.80 & 7058.83 & 0.259 \\
2455949.7582 &   5.07 & 1.85 &  -8.51 & 7053.47 & 0.267 \\
2455950.5623 & -15.54 & 1.31 &  -9.51 & 7052.13 & 0.271 \\
2455950.6682 & -14.64 & 1.11 & -23.11 & 7051.31 & 0.267 \\
2455950.7686 & -19.08 & 1.44 & -19.07 & 7046.46 & 0.257 \\
2455951.5488 & -12.78 & 1.57 &  -8.83 & 7025.98 & 0.264 \\
2455951.6558 & -16.45 & 1.87 &  -5.73 & 7029.55 & 0.263 \\
2455951.7570 & -18.19 & 1.90 &  -9.35 & 7041.75 & 0.239 \\
2455952.5652 & -12.29 & 1.78 & -25.85 & 7031.26 & 0.264 \\
2455952.7702 &  -7.99 & 1.59 &  -9.25 & 7039.82 & 0.271 \\
2455953.5560 & -17.12 & 1.48 & -32.88 & 7023.51 & 0.288 \\
2455953.6847 & -14.07 & 1.57 & -33.36 & 7044.27 & 0.273 \\
2455953.7630 &  -6.41 & 1.76 &  -8.40 & 7041.29 & 0.266 \\
2455954.5540 & -12.13 & 1.38 & -28.15 & 7039.45 & 0.275 \\
2455954.6379 &  -9.08 & 1.59 & -27.34 & 7054.85 & 0.284 \\
2455955.5585 &  -0.98 & 1.76 & -26.23 & 7049.09 & 0.264 \\
2455955.6389 &   0.00 & 1.56 & -24.63 & 7054.10 & 0.269 \\
2455955.7328 &   0.53 & 1.37 & -28.13 & 7037.54 & 0.273 \\
2455956.6246 &   5.90 & 1.23 & -16.30 & 7059.86 & 0.284 \\
2455956.7290 &   1.62 & 1.45 & -18.90 & 7058.08 & 0.293 \\
2455957.6437 &  -4.87 & 1.60 & -12.78 & 7066.72 & 0.264 \\
2455958.5668 &  -8.89 & 1.70 & -14.47 & 7067.97 & 0.268 \\
2455958.6585 &  -6.76 & 1.56 & -18.02 & 7054.47 & 0.264 \\
2455958.7173 &  -5.81 & 1.45 &  -6.45 & 7060.02 & 0.301 \\
2455959.5536 &  -3.51 & 1.78 & -30.62 & 7066.75 & 0.283 \\
2455959.6410 &   2.06 & 1.75 & -22.66 & 7063.47 & 0.285 \\
2455959.7221 &   0.15 & 1.61 & -18.01 & 7048.00 & 0.286 \\
2455960.5499 &  -1.46 & 1.67 & -11.92 & 7072.93 & 0.290 \\
2455960.6422 &  -3.96 & 1.47 & -18.31 & 7069.42 & 0.296 \\
2455960.7181 &  -2.78 & 1.61 & -12.69 & 7063.44 & 0.300 \\
2455961.5713 & -13.67 & 1.74 & -22.78 & 7061.55 & 0.276 \\
2455961.7114 & -12.47 & 1.53 & -19.50 & 7063.92 & 0.277 \\
2455962.5420 &  -7.57 & 1.97 & -32.92 & 7077.07 & 0.278 \\
2455962.6334 &  -7.67 & 1.69 & -22.88 & 7072.07 & 0.276 \\
2455962.7231 & -12.91 & 2.39 & -25.95 & 7064.79 & 0.276 \\
2455963.5585 &  -4.10 & 1.69 & -26.40 & 7067.31 & 0.288 \\
2455963.6485 &  -6.18 & 1.51 & -20.01 & 7070.92 & 0.305 \\
2455963.7044 &  -5.60 & 1.54 & -18.69 & 7069.43 & 0.290 \\
2455964.5581 &  -9.83 & 1.77 & -10.56 & 7074.88 & 0.280 \\
2455964.6253 &  -7.27 & 1.83 & -16.87 & 7071.36 & 0.271 \\
2455964.7036 & -14.89 & 2.24 & -12.06 & 7069.06 & 0.279 \\
\hline \hline
\end{tabular}
\end{table*}

\label{lastpage}

\end{document}